\documentstyle[preprint,prl,aps]{revtex}
\tightenlines
\begin{document}
%
%
\draft
\title{Detecting new physics contributions to the
$D^0 - \overline{D^0}$ mixing\\
through their effects on $B$ decays}
\author{Carlos C.\ Meca$^1$ and Jo{\~a}o P.\ Silva$^{1,2}$}
\address{$^1$ Centro de F{\'\i}sica Nuclear da Univ. de Lisboa\\
Av.\ Prof.\ Gama Pinto, 2,\\
1699 Lisboa Codex, Portugal}
\address{$^2$ Centro de F{\'\i}sica\\
Instituto Superior de Engenharia de Lisboa,\\
1900 Lisboa, Portugal}
\date{\today}
\maketitle
\begin{abstract}
New physics effects may yield a detectable mass difference in the
$D^0 - \overline{D^0}$ system, $\Delta m_D$.
Here we show that this has an important impact on some
$B \rightarrow D$ decays.
The effect involves a new source of CP violation,
which arises from the interference between the phases in the $B \rightarrow D$
decays and those in the $D^0 - \overline{D^0}$ system.
This interference is naturally large.
New physics may well manifest itself through $\Delta m_D$ contributions
to these $B$ decays.
\end{abstract}
11.30.Er  12.60.-i  13.25.Hw  14.40.Lb  

\narrowtext

The $D^0 - \overline{D^0}$ mixing is expected to be very small in the
standard model (SM).
Although estimates of the long-distance contributions to
$x_D \equiv \Delta m_D/\Gamma_D$ are rather uncertain
\cite{Wol85,Don86,Geo92,Ohl93,Gol98},
it is clear that,
in the SM,
$x_D$ should be orders of magnitude smaller than the
current experimental limit of $|x_D| < 0.1$ \cite{PDG}.
On the other hand,
values of $x_D \sim 10^{-2}$ are easily obtained in
many models of new physics \cite{several}.

It has been stressed that an effect of this magnitude
can be detected by tracing the time dependence of the decays
of neutral $D$ mesons \cite{Liu94}.
The idea is that, in the presence of mixing,
an initial $D^0$ meson will evolve in time according to
\begin{equation}
e^{\tau/2} D^0(\tau)
= \left( 1 - \frac{x_D^2 \tau^2}{8} \right) D^0
- \frac{i}{2} \frac{q_D}{p_D} x_D \tau  \overline{D^0},
\end{equation}
where $D_{H,L} = p_D D^0 \pm q_D \overline{D^0}$,
$\tau = \Gamma_D t$,
and we have only keep terms up to order $(x_D \tau)^2$.
Therefore,
the time-dependent decay rate will depart from the usual exponential
falloff, $e^{- \tau}$,
with a leading correction proportional to $x_D \tau e^{- \tau}$.

In this letter,
we wish to show that there is a corresponding effect in the
decays of $B$ mesons which involve neutral $D$ mesons as an intermediate state.
To be specific \cite{upcoming},
we will concentrate on
$B^\pm \rightarrow K^\pm D \rightarrow K^\pm f_D$ decay chains,
where $D$ refers to an intermediate $D^0 - \overline{D^0}$ combination,
which then decays into the final state $f_D$.
We will show that:
\begin{itemize}
\item one may, in principle, use the
$B^\pm \rightarrow K^\pm D \rightarrow K^\pm (X l \nu)_D$ decay chain,
to get at $x_D$.
Here,
$(X l \nu)_D$ refers to the state that is obtained from the
semileptonic decay of the neutral $D$ meson;
\item the previous decay chain involves a novel rephasing-invariant
parameter, which we denote by $\xi_i$,
and which beats the phase in the amplitude of the $B^\pm$
decay {\it into} the neutral $D$ meson,
against the phase in the $D^0 - \overline{D^0}$ mixing.
This is different from the usual $\lambda_f$ parameters which beat the
phase in $D^0 - \overline{D^0}$ mixing against the phase in the amplitude
of the decay {\it from} the $D^0 - \overline{D^0}$ state into a final state
$f_D$.
The crucial difference is that,
while the interference CP violation in $D$ decays
(probed by $\mbox{Im} \lambda_f$) is small in the SM,
the novel interference CP violation described here
(and probed by $\mbox{Im} \xi_i$)
in necessarily large, even within the SM.
If the CP-violation in the $D^0 - \overline{D^0}$ mixing
turns out to be very small,
a large value for $x_D$ might be detected in the decay chains discussed
here,
before it can be probed in the decays of neutral $D$ mesons.
\item this new physics effect may also have a considerable impact on
the method proposed by Atwood, Dunietz and Soni (ADS)
to get at the CKM phase $\gamma$ through the decay chain
$B^\pm \rightarrow K^\pm D \rightarrow K^\pm f_D$ \cite{ADS},
should $x_D$ be of order $10^{-2}$.
\end{itemize}

The Cabibbo allowed $\overline{D^0} \rightarrow K^+ \pi^-$ decay amplitude 
and the doubly Cabibbo suppressed decay amplitude (DCSD) for
$D^0 \rightarrow K^+ \pi^-$, may be parametrized as
\begin{eqnarray}
A_{\overline{D^0} \rightarrow K^+ \pi^-}
= A_{D^0 \rightarrow K^- \pi^+}
&=&
A\, e^{i \bar{\delta}_D}.
\nonumber\\
A_{D^0 \rightarrow K^+ \pi^-}
= A_{\overline{D^0} \rightarrow K^- \pi^+}
&=&
- \epsilon \,A\, e^{i \delta_D}
\nonumber
\end{eqnarray}
In writing these expressions we have neglected 
the phase $\arg(- V_{cd} V_{us} V_{cs}^\ast V_{ud}^\ast)$,
which lies around 0.003 radians in the SM \cite{Ale94},
and is likely to remain small in many of the models of new physics.
An estimate based on the tree-level diagrams,
would lead to
$\epsilon \sim |(V_{cd} V_{us})/(V_{cs} V_{ud})| \sim 0.05$,
but this estimate can be affected by form-factor and decay constants.
To overcome this problem,
Atwood, Dunietz and Soni \cite{ADS} use instead the central value of the
experimental ratio of branching ratios determined by CLEO \cite{CLEO94}
to get
$ \epsilon^2 \sim
\mbox{BR}[D^0 \rightarrow K^+ \pi^-]/
\mbox{BR}[\overline{D^0} \rightarrow K^+ \pi^-]
\sim (0.088)^2$.
However,
this assumes that $x_D = 0$,
which is precisely the assumption that we wish to remove.
In any case,
the exact value of this parameter does not affect our argument.

The decay rate for $D^0 \rightarrow K^+ \pi^-$ is determined by
the rephasing invariant parameter
\[
\lambda_{K^+ \pi^-} \equiv \frac{q_D}{p_D}
\frac{A_{\overline{D^0} \rightarrow K^+ \pi^-}}{A_{D^0 \rightarrow K^+\pi^-}}
= - \frac{1}{\epsilon} e^{i(2 \theta_D - \Delta_D)},
\]
where we have used $q_D/p_D \equiv e ^{2 i \theta_D}$,
and $\Delta_D \equiv \delta_D - \bar{\delta}_D$.
In fact,
\begin{eqnarray}
e^{\tau}\, R[D^0 \rightarrow K^+ \pi^-] 
&\sim&
\left| A_{D^0 \rightarrow K^+\pi^-} \right|^2
\left[
1 + \mbox{Im} \lambda_{K^+ \pi^-}\, x_D \tau
+
\left(|\lambda_{K^+ \pi^-}|^2 - 1 \right)
\left( \frac{x_D \tau}{2} \right)^2
\right]
\nonumber\\
&\sim&
\epsilon^2 A^2
\left[ 1 - \frac{x_D \tau}{\epsilon}
\sin{(2 \theta_D - \Delta_D)}
+ \left( \frac{x_D \tau}{2 \epsilon} \right)^2
\right].
\label{Wolf}
\end{eqnarray}
In the SM,
and with the usual phase conventions,
the $D^0 - \overline{D^0}$ mixing phase, $\theta_D$,
is very small.
Then,
the term linear in $x_D \tau$ can only arise due to
to the CP-conserving strong phase difference $\Delta_D$
\cite{Bla95,Bro95}.
But,
as Wolfenstein has pointed out,
$\theta_D$ may be large in many of the models for which 
$x_D \sim 10^{-2}$ is allowed \cite{Wol95}.
In that case,
the linear term is due to the CP violation in the interference
between the decay and the $D^0 - \overline{D^0}$ mixing.
Wolfenstein then makes the important point that,
since we are looking at a DCSD,
the linear term is enhanced by $1/\epsilon$.
As a result,
for $x_D \tau = 0.02$,
the linear term may affect the decay rate by up to $40\%$,
while the quadratic term only shows up at the $4\%$ level
\cite{Wol95}.

Our first observation is that a similar effect arises in the
$B^\pm \rightarrow K^\pm D \rightarrow K^\pm (X l \nu)_D$ decay chain.
We parametrize,
\begin{eqnarray}
A_{B^+ \rightarrow K^+ \overline{D^0}}
=
B\, e^{i \bar{\delta}_B}\ ,
& \ \ \ &
A_{B^+ \rightarrow K^+ D^0}
=
\tilde{\epsilon}\,B\, e^{i\gamma} e^{i \delta_B}\ ,
\nonumber\\ 
A_{B^- \rightarrow K^- D^0}
=
B\, e^{i \bar{\delta}_B}\ ,
& \ \ \ &
A_{B^- \rightarrow K^- \overline{D^0}}
=
\tilde{\epsilon}\,B\, e^{-i\gamma} e^{i \delta_B}\ .
\nonumber
\end{eqnarray}
Here,
we use
$\gamma \equiv \arg(- V_{ud} V_{cb} V_{ub}^\ast V_{cd}^\ast)$,
and we continue to neglect
$\arg(- V_{cd} V_{us} V_{cs}^\ast V_{ud}^\ast)$.
We will take the estimate
$\tilde{\epsilon} \sim |(V_{ub} V_{cs}^\ast)/(V_{cb} V_{us}^\ast)| |a_2/a_1|
\sim 0.09$ \cite{ADS},
where $|V_{ub}/V_{cb}| \sim 0.08$ has been used,
and the phenomenological $|a_2/a_1| \sim 0.26$ factor accounts for the fact
that the $B^+ \rightarrow K^+ D^0$ decay is color-suppressed,
while the $B^+ \rightarrow K^+ \overline{D^0}$ decay is not.
As happens with the estimate of $\epsilon$,
this estimate of $\tilde{\epsilon}$ is subject to uncertainties due to
final state interactions combined with flavour-SU(3) breaking effects.
This does not affect the point we wish to make,
although it will affect the precise value of the effects to be discussed.
The relevant decay rates will depend on the rephasing-invariant
parameter \cite{upcoming}
\[
\xi_+ \equiv 
\frac{A_{B^+ \rightarrow K^+ \overline{D^0}}}{
A_{B^+ \rightarrow K^+ D^0}} \ \frac{p_D}{q_D}=
\frac{1}{\tilde{\epsilon}} e^{- i \gamma} e^{-i(2 \theta_D + \Delta_B)},
\]
where $\Delta_B \equiv \delta_B - \bar{\delta}_B$.
Indeed
\begin{eqnarray}
e^{\tau}\, 
R{\left[B^+ \rightarrow K^+ (X^- l^+ \nu_l)_D\right]}
& \sim &
\left| A_{D^0 \rightarrow l^+} \right|^2
\left| A_{B^+ \rightarrow K^+ D^0} \right|^2
\left[
1 + \mbox{Im} \xi_+\, x_D \tau
+
\left(|\xi_+|^2 - 1 \right) \left( \frac{x_D \tau}{2} \right)^2
\right]
\nonumber\\
& \sim &
\left| A_{D^0 \rightarrow l^+} \right|^2
\tilde{\epsilon}^2 B^2
\left[ 1 - \frac{x_D \tau}{\tilde{\epsilon}}
\sin{(\gamma + 2 \theta_D + \Delta_B)}
+ \left( \frac{x_D \tau}{2 \tilde{\epsilon}} \right)^2
\right].
\label{Silva}
\end{eqnarray}
A similar analysis shows that,
for the CP conjugated process,
the decay rate has the same expression,
except for changes in the signs multiplying $x_D$ and $\Delta_B$.
Eq.\ (\ref{Silva}) should be compared with Eq.\ (\ref{Wolf}).
In particular, one should notice the same $1/\tilde{\epsilon}$-type
enhancement of the linear term.
This occurs because the unmixed decay path,
$B^+ \rightarrow K^+ D^0 \rightarrow K^+ (X^- l^+ \nu_l)_D$, 
involves the suppressed decay $B^+ \rightarrow K^+ D^0$,
while the mixed decay path,
$B^+ \rightarrow K^+ \overline{D^0} \rightarrow K^+ (X^- l^+ \nu_l)_D$,
involves the much larger $B^+ \rightarrow K^+ \overline{D^0}$ decay amplitude.
As before,
the linear term may be detected,
should the weak and/or strong phase differences be large.
The crucial point is that, here, the weak phase is necessarily large,
for it involves the phase $\gamma$ present in the $B \rightarrow K D$ decays
\cite{Xin96}.
Therefore,
even if $\theta_D$ turns out to be small,
a value of $x_D \sim 10^{-2}$ will have a large impact on decay chains of the
type $B^\pm \rightarrow D K^\pm \rightarrow f_D K^\pm$.

This effect might be easiest to detect through the time-integrated,
CP-violating asymmetry,
\begin{eqnarray}
& &
\frac{\Gamma{\left[B^- \rightarrow K^- (X^+ l^- \bar{\nu}_l)_D\right]} -
\Gamma{\left[B^+ \rightarrow K^+ (X^- l^+ \nu_l)_D\right]}}{
\Gamma{\left[B^- \rightarrow K^- (X^+ l^- \bar{\nu}_l)_D\right]}
+\Gamma{\left[B^+ \rightarrow K^+ (X^- l^+ \nu_l)_D\right]}}
\nonumber\\
& &
=
\frac{x_D}{\tilde{\epsilon}}
\frac{\sin{(\gamma + 2 \theta_D)} \cos{\Delta_B}}{1
- \cos{(\gamma + 2 \theta_D)} \sin{\Delta_B} \frac{x_D}{\tilde{\epsilon}}},
\nonumber
\end{eqnarray}
where we have only taken the expansions up to terms of order $x_D$.
For $x_D \sim 10^{-2}$,
this asymmetry is naturally of order $10\%$.

However,
the experiment described thus far might be difficult to implement,
since it is subject to large backgrounds.
Notably,
from the direct semileptonic decay $B^+ \rightarrow X l^+ \nu_l$.
This background may be reduced by using a number of topological and
kinematical properties that distinguish the signal \cite{ADS}.
In the SM,
$x_D$ is very small,
and the decay chain in Eq.\ (\ref{Silva}) yields the branching ratio for
$B^+ \rightarrow K^+ D^0$.
This observable is needed for the 
Gronau--London--Wyler method to measure $\gamma$ \cite{GL,GW},
which consists in comparing the $B^+ \rightarrow K^+ D^0$ and 
$B^+ \rightarrow K^+ \overline{D^0}$ decays with
$B^+ \rightarrow K^+ D_{\rm cp}$ \cite{DCP}.
Thus,
this method is impaired by the same experimental difficulties.
What we have shown is that,
should the experiment turn out to be feasible,
the decay rate (and, thus, the Gronau--London--Wyler method)
will be affected by $x_D \sim 10^{-2}$,
at a detectable level.

The effect found in Eq.\ (\ref{Silva}) also plays an important role in the
Atwood--Dunietz--Soni (ADS) method to get at the phase $\gamma$.
That method is based upon the decay chain
\[
A_{\rm ADS}^0 \equiv 
A_{B^+ \rightarrow K^+ D^0} A_{D^0 \rightarrow f_D} +
A_{B^+ \rightarrow K^+ \overline{D^0}} A_{\overline{D^0} \rightarrow f_D}
\]
One uses two final states $f_D$ for which
$\overline{D^0} \rightarrow f_D$ is a DCSD,
while $D^0 \rightarrow f_D$ is Cabibbo allowed,
and the CP-conjugated decays.
Examples include $f_D = K^- \pi^+$, $K^- \rho^+$, $K \pi \pi$, etc.
The idea is to profit from the fact that,
being both suppressed (by $\tilde{\epsilon}$ and $\epsilon$,
respectively),
the two decay paths interfere at $O(1)$.
For example,
taking $f_D = K^- \pi^+$,
one gets
\begin{equation}
|A_{\rm ADS}^0|^2 =
A^2 B^2
\left[
\tilde{\epsilon}^2
+ \epsilon^2
- 2 \tilde{\epsilon} \epsilon \cos{(\gamma + \Delta_B - \Delta_D)}
\right]
\label{AADS^2}
\end{equation}
The fact that the interference is of $O(1)$ has three important consequences.
First,
it makes it difficult to use this decay chain to determine the
$\tilde{\epsilon}$-suppressed $B^+ \rightarrow K^+ D^0$ decay amplitude,
which is needed for the
the Gronau--London--Wyler method
to measure $\gamma$ \cite{GL,GW}.
Second,
it allows one to measure $\gamma$ without the need to measure
$B^+ \rightarrow K^+ D^0$,
by using two such decays and their CP conjugates.
For example,
one would look for the decays
$B^+ \rightarrow K^+ (K^+ \pi^-)_D$,
$B^+ \rightarrow K^+ (K^+ \rho^-)_D$,
$B^- \rightarrow K^- (K^- \pi^+)_D$,
and
$B^- \rightarrow K^- (K^- \rho^+)_D$.
These four decays are fitted for the four unknowns:
the branching ratio for $B^+ \rightarrow K^+ D^0$,
two strong phase factors (one for each $f_D$) and the weak phase $\gamma$.
This is the method proposed by Atwood, Dunietz and Soni
\cite{ADS}.
Third,
it makes it likely for this decay chain to be affected by terms
proportional to $x_D$,
corresponding to amplitudes that are not CKM or color suppressed.

This is indeed the case.
To lowest order in $x_D \tau$,
the corrections arise from
\[
A_{B^+ \rightarrow K^+ \overline{D^0}}\ \frac{p_D}{q_D}
\ A_{D^0 \rightarrow f_D}
+
A_{B^+ \rightarrow K^+ D^0}\  \frac{q_D}{p_D}
\ A_{\overline{D^0} \rightarrow f_D}.
\]
The first decay path does not have any $\epsilon$ or $\tilde{\epsilon}$
suppression.
Thus,
the corresponding amplitude exhibits the required $1/\epsilon$-type
enhancement with respect to the usual amplitude $A_{\rm ADS}^0$.
This is analogous to the linear terms discussed in
Eqs.\ (\ref{Wolf}) and (\ref{Silva}).
As a result,
the time-integrated decay rate which appears in the ADS method
is no longer given solely by $|A_{\rm ADS}^0|^2$.
To leading order in $x_D$,
one must add to $|A_{\rm ADS}^0|^2$ in Eq.\ (\ref{AADS^2}),
a correction given by
\begin{equation}
x_D A^2 B^2 
\left[
- \tilde{\epsilon} \sin{(\gamma + 2 \theta_D + \Delta_B)}
+ \epsilon \sin{(2 \theta_D + \Delta_D)}
\right],
\label{Silva-ADS}
\end{equation}
where we have neglected the highly suppressed decay sequence in which
$B^+ \rightarrow K^+ D^0$ is followed by
$\overline{D^0} \rightarrow f_D$.
The first term in Eq.\ (\ref{Silva-ADS})
has the same origin of the linear term in Eq.\ (\ref{Silva});
its correction is large even if the CP violation in the
$D^0 - \overline{D^0}$ system remains negligible.
The second term in Eq.\ (\ref{Silva-ADS})
is similar to that in Eq.\ (\ref{Wolf});
it gives an important correction if the phase $\theta_D$ is sizeable. 
In any case, barring large cancellations,
the correction is of order $10\%$ for $x_D \sim 10^{-2}$.
This highlights the discovery potential
that the $B^\pm \rightarrow K^\pm D \rightarrow K^\pm f_D$ decay chains
have to probe new physics effects in the $D^0 - \overline{D^0}$ system.
Conversely,
it shows that such new physics effects may have a considerable
impact on methods probing the angle $\gamma$,
such as the ADS method.

A final note should be made about the relevance of the $\xi_i$
parameters.
Usually,
one classifies the CP violation present in decays {\it from} neutral meson
systems in terms of the CP violation in the mixing,
CP violation in the decay amplitudes,
and CP violation in the interference between the mixing and the
decay amplitudes
{\it from} that neutral meson system into the final state.
For example, in the decays of $D^0$ into a CP eigenstate $f$,
these tree types of CP violation are measured by
$|q_D/p_D|-1$,
$|A_{D^0 \rightarrow f}/A_{\overline{D^0} \rightarrow f}|-1$,
and $\mbox{Im} \lambda_f$,
respectively.
But,
when discussing decay chains with neutral meson systems in an intermediate
state,
a new type of CP-violation may be possible;
that arising from the interference between the mixing and the decay amplitudes
from the initial state {\it into} that neutral meson system.
This effect does not show up in the usual discussions of the
ADS and related methods because it is assumed that,
in accordance with the SM,
the neutral $D$ mesons do not exhibit oscillations at a detectable level.
This discussion can also be avoided in the $B \rightarrow K$ cascade decays
discussed by Azimov \cite{cas-Azi}, Kayser \cite{cas-Kay}
and others \cite{cas-others}.
The reason is that,
in those cases,
one only has {\it one} decay amplitude from a given initial $B$ meson
flavor {\it into} the neutral kaon system.
For instance,
the $B^0_d \rightarrow J/\psi \overline{K^0}$ does not occur in the SM.
Here,
the presence of two decay paths,
which is generic in $B \rightarrow D$ and $D \rightarrow K$ transitions,
forced us into this new type of CP violation \cite{Xin96}.
The first term in Eq.\ (\ref{Silva-ADS}) comes from $\mbox{Im} \xi_+$,
and involves the phase $\gamma$ from the $B \rightarrow D$ decays;
the second term comes from $\mbox{Im} \lambda_{K^- \pi^+}$
and involves only the usual phases in $D$ decays.
It can be shown that all cascade decays,
and the interference CP violations present therein,
may be parametrized in terms of the $\xi_i$ and $\lambda_f$ parameters
alone \cite{upcoming}.

In summary,
we have seen that the suggestion by Wolfenstein \cite{Wol95} that
values of $x_D \sim 10^{-2}$ might be easier to detect in the 
CP violating terms of $D \rightarrow f$ decays,
is even more important for $B \rightarrow D$ decays.
Here,
the relevant phase involves $\gamma$ and is naturally large,
even if the new phase in $D^0 - \overline{D^0}$ mixing turns out
to be very small.
Moreover,
this effect has a strong impact on the
$B^\pm \rightarrow K^\pm D \rightarrow K^\pm f_D$ decay chains,
which are at the heart of the ADS method to measure $\gamma$
\cite{ADS}.
When discussing how new physics affects these measurements,
one cannot concentrate exclusively on new physics in the $B$ system
since, as shown here, new physics effects in the $D$ system are likely to
play a crucial role as well.
Finally,
we note that the interference effects described by $\xi_i$
are every bit as important for these decay chains,
as the celebrated interference effects contained in the 
$\lambda_f$ parameters.

 
\vspace{5mm}
 
We would like to thank L.\ Wolfenstein for reading and criticizing
this manuscript.
We also thank A.\ Amorim and M.\ Santos for this and for a fruitful
and enjoyable collaboration on related subjects.
We are especially indebted to B.\ Kayser for the excellent series 
of tutorials on cascade decays that he gave at the
Centro de F{\'\i}sica Nuclear da Univ. de Lisboa.
The work of J.\ P.\ S.\ was supported in part by the portuguese
JNICT under contract CERN/P/FIS/1096/96.


%
%
%


\begin{references}
\bibitem{DCP}
By $D_{\rm cp}$ one means that the intermediate $D^0 - \overline{D^0}$
combination has been detected through its decay into a CP eigenstate.
\end{references}
\end{document}